\documentclass[journal]{IEEEtran}
\ifCLASSINFOpdf
\usepackage[pdftex]{graphicx}
\else
\fi
\usepackage{amssymb, amsmath}
\usepackage{algorithm, algorithmic}

\usepackage{array}
\usepackage{multirow}
\usepackage{url}

\usepackage{xcolor}

\begin{document}
  %
  \title{Multi-agent Reinforcement Learning-based Joint Design of Low-Carbon P2P
  Market and Bidding Strategy in Microgrids}
  %
  %
  %

  \author{Junhao Ren, Honglin Gao, Sijie Wang, Lan Zhao, Qiyu Kang, Aniq Ashan, Yajuan Sun,
  Gaoxi, Xiao,~\IEEEmembership{Senior Member,~IEEE}
  \thanks{Junhao Ren, Honglin Gao, Sijie Wang, Lan Zhao, Gaoxi Xiao are with the
  School of Electrical and Electronic Engineering, Nanyang Technological University,
  639798, Singapore, E-mail: (junhao002@e.ntu.edu.sg, honglin001@e.ntu.edu.sg,
  wang1679@e.ntu.edu.sg, zhao0468@e.ntu.edu.sg, egxxiao@ntu.edu.sg).}
  \thanks{Qiyu Kang is with the School of Information Science and Technology, University
  of Science and Technology of China, Hefei, 230026, China, E-mail:
  qiyukang@ustc.edu.cn.}
  \thanks{Aniq Ashan is with the MOYA Analytics, 051912, Singapore, E-mail: aniq@moya-analytics.com.}
  \thanks{Yajuan Sun is with the Singapore Institute of Manufacturing Technology,
  Agency for Science, Technology and Research, 636732, Singapore, E-mail: sun\_yajuan@simtech.a-star.edu.sg.}
  }

  \markboth{Journal of \LaTeX\ Class Files,~Vol.~14, No.~8, December~2025}%
  {Shell \MakeLowercase{\textit{et al.}}: Bare Demo of IEEEtran.cls for IEEE Journals}
  %



  \maketitle

\begin{abstract}
The challenges of the uncertainties in renewable energy generation and the instability of the real-time market limit the effective utilization of clean energy in microgrid communities. Existing peer-to-peer (P2P) and microgrid coordination approaches typically rely on certain centralized optimization or restrictive coordination rules which are difficult to be implemented in real-life applications. To address the challenge, we propose an intraday P2P trading framework that allows self-interested microgrids to pursue their economic benefits, while allowing the market operator to maximize the social welfare, namely the low carbon emission objective, of the entire community. Specifically, the decision-making processes of the microgrids are formulated as a Decentralized Partially Observable Markov Decision Process (DEC-POMDP) and solved using a Multi-Agent Reinforcement Learning (MARL) framework. Such an approach grants each microgrid a high degree of decision-making autonomy, while a novel market clearing mechanism is introduced to provide macro-regulation, incentivizing microgrids to prioritize local renewable energy consumption and hence reduce carbon emissions. Simulation results demonstrate that the combination of the self-interested bidding strategy and the P2P market design helps significantly improve renewable energy utilization and reduce reliance on external electricity with high carbon-emissions. The framework achieves a balanced  integration of local autonomy, self-interest pursuit, and improved community-level economic and environmental benefits.
\end{abstract}

\begin{IEEEkeywords}
Multi-microgrid systems, Peer-to-peer market, multi-agent reinforcement learning, low-carbon emission.
\end{IEEEkeywords}

\IEEEpeerreviewmaketitle

\section{Introduction}
Carbon emissions, the primary driver of the greenhouse effect, have severely constrained
  global sustainable development. To address this challenge, the Paris Agreement
  \cite{horowitz2016paris} established the critical goal of achieving net-zero emissions
  by 2050. Among various mitigation strategies, the large-scale integration of
  renewable energy sources (RES) into conventional power networks has emerged as
  an approach of fundamental importance \cite{heRapidCostDecrease2020}. It is projected that renewable sources will account for $ 25 \% - 41 \% $ of global electricity generation by 2040 \cite{dengPowerSystemPlanning2020}. Such a transition however marks a paradigm shift
  from centralized generation to distributed energy resources (DERs) \cite{alanne2006distributed}, fundamentally altering the operation of power systems.

  The high penetration of renewables introduces significant operational risks
  due to their inherent stochasticity and intermittency. Such characteristics may
  lead to severe power fluctuations that challenge existing demand–response mechanisms.
  While Energy Storage Systems (ESS) are effective in mitigating these
  fluctuations \cite{kwonOptimalDayAheadPower2017}, their high deployment costs remain
  a barrier. At the same time, traditional consumers are evolving into ``prosumers"
  capable of both generation and consumption. For microgrids with limited
  storage or generation capacities, relying solely on the main grid often proves economically
  inefficient due to widening spreads between buying and selling prices.
  Consequently, Peer-to-Peer (P2P) energy trading has emerged as a promising solution
  \cite{morstynUsingPeertopeerEnergytrading2018, liuComparisonCentralizedPeertoPeer2022}.
  By enabling direct energy exchange between neighbors, P2P markets facilitate energy
  management, reduce transmission losses, and improve utilization of locally generated renewable energy \cite{morstynUsingPeertopeerEnergytrading2018}.

 Despite the benefits of P2P markets, their practical implementation faces
 complex challenges. The trading environment is highly dynamic, characterized
 by uncertainties in renewable generation and conflicting interests of
 different participants. A natural baseline is centralized optimization;
 however, centralized approaches often struggle with computational scalability
 and require private user data, which raises privacy and trust concerns \cite{liuComparisonCentralizedPeertoPeer2022, luoOptimizationSolarbasedIntegrated2021}. 
 These limitations motivate {\it distributed} P2P market clearing, where
 participants solve local subproblems and coordinate through limited
 information exchange. For example,
 \cite{wangStochasticCooperativeBidding2022} formulates a stochastic cartel
 game and applies diagonal quadratic approximation to decompose the problem
 into microgrid-level local subproblems, while
 \cite{luPeertopeerJointElectricity2023} models joint electricity--carbon
 trading as a Nash--Stackelberg bi-level game and uses ADMM to simplify the
 lower-level distributed P2P trading problem.

Several challenges need to be handled in a P2P electricity market with traditional decentralized optimization methods. First, many existing distributed optimization methods are built upon cooperative (or coalition-based) game formulations, whereas practical P2P participants are often self-interested and may behave strategically. Second, the decision space and decision process of each microgrid can be overly constrained by rigid bidding/clearing mechanisms: the bidding variables are often limited to prices, while other private information is treated as exogenous and truthfully reported, which restricts autonomy in local energy management and strategic trading. Third, carbon-emission considerations are often imposed as hard constraints rather than being encoded into market incentives, providing only limited economic motivation for proactive decarbonization \cite{xuGameBasedPricingJoint2024}.
  
Multi-Agent Reinforcement Learning (MARL) offers a favorable option for
 tackling the challenges mentioned above \cite{yu_surprising_2022}. Specifically,
 MARL allows distributed energy entities to make autonomous and adaptive
 decisions under partial observability. By allowing microgrids to learn through
 continuous interactions, MARL helps pave the way toward self-organized
 intelligent energy systems capable of robustly mitigating the operational
instability and uncertainties arising from renewable intermittency and prediction errors, and steadily achieve low carbon emission of the whole system.

 In this study, we propose a decentralized market-learning framework for real-time P2P electricity trading of microgrids. The core principle of this framework is to couple an incentive-aware market clearing mechanism with decentralized sequential decision-making: the market mechanism encodes system-level objectives into bidding clearing mechanism, while microgrids independently adapt their strategies based on local observations and historical interactions. In this framework, coordinated system-level outcomes are not imposed explicitly but emerge from the interaction between rational agents and the designed market incentives, making it suitable for complex, partially observable P2P environments.
  The main contributions of this paper are summarized as follows:
  \begin{itemize}
    \item \textbf{Non-Cooperative Game Formulation.} We explicitly model microgrids as rational and self-interested agents, departing from cooperative or truth-telling assumptions commonly adopted in existing P2P market studies. Their interactions are formulated as a  DEC-POMDP (Decentralized Partially Observable Markov Decision Process) enabled, MARL-based learning of strategic behaviors under partial observability.

    \item \textbf{Joint Bidding and Energy Storage Decision Space.} The proposed framework endows microgrids with a coupled and flexible decision space, allowing them to jointly determine trading roles, bidding prices, traded quantities, and energy storage schedules. This design captures the intrinsic coupling between market participation and internal energy management that is overlooked by price-only or fixed-role models.

    \item \textbf{Carbon-Aware Incentive Design.} The proposed framework incorporates carbon emission considerations into both the pricing signals and the double-auction clearing process, rather than imposing rigid emission constraints. By jointly shaping individual payoffs and trading priorities, the proposed design economically incentivizes low-carbon, locally balanced trading behaviors while preserving decentralized decision autonomy.
  \end{itemize}
  The remainder of the paper is organized as follows. Section II reviews
  existing methods and technologies employed in the P2P market of multi-microgrid
  systems. Section III presents the overall system formulation, including the
  structure of the multi-microgrid systems and the electricity market framework.
  Section IV describes the proposed MARL-based model for the P2P market. Section
  V provides simulation results and discussions. Section VI concludes the paper and
  outlines potential directions for future research.

  \section{Related Works}
  Existing studies on trading problems in multi-microgrid systems mainly rely on
  two methodological paradigms: model-based optimization and model-free
  optimization (e.g., MARL).

  Model-based optimization refers to frameworks in which the objective functions,
  constraints, and system dynamics are known or can be analytically modeled,
  allowing the optimal decisions to be directly derived from the model. In \cite{liRiskaverseEnergyTrading2017},
  a two-stage stochastic game model was proposed to determine the optimal energy
  trading strategy under supply and demand uncertainties. A distributed algorithm
  was further developed to obtain the stochastic Nash equilibrium using the sample
  average approximation technique. A stochastic cartel game was introduced in
  \cite{wangStochasticCooperativeBidding2022} to address cooperative bidding
  strategies among multiple microgrids in P2P energy transactions considering
  renewable generation and energy storage. A collaborative optimization problem for
  capacity planning of distributed generation units and P2P trading was investigated
  in \cite{wangRiskaverseStochasticCapacity2024}, where a risk-averse stochastic
  programming framework and Nash bargaining approach were adopted to mitigate
  renewable energy uncertainties and ensure fairness in trading. Further, a
  low-carbon P2P trading model based on a master–slave nested mixed game was
  proposed in \cite{liangMultiagentLowcarbonOptimal2024}, which explicitly incorporated
  carbon emission flows as constraints in the game formulation.

  Model-free optimization, on the other hand, enables agents to learn their individual
  trading strategies and energy schedules from historical data to maximize operational
  profitability, or optimize a certain other objective function, without requiring explicit knowledge of system dynamics. In
  \cite{qiuMultiagentReinforcementLearning2021}, a double-sided auction-based P2P
  electricity market was formulated as a MARL problem, and a DA-MADDPG
  (Double-Side Auction Multi-Agent Deep Deterministic Policy Gradient) algorithm was proposed to maximize the profits of prosumers in a dynamic
  electricity market. A two-level MARL framework was developed in \cite{mayMultiagentReinforcementLearning2023}
  for designing dynamic pricing policies that facilitate efficient on-site energy trading
  while supporting decarbonization and grid security objectives. In
  \cite{yangMultistageStochasticDispatching2024}, a multi-stage dispatching method
  incorporating day-ahead and intraday scheduling was proposed to address uncertainties
  in electricity–hydrogen integrated energy systems. Moreover, \cite{chenCombinedCarbonCapture2024}
  introduced a multi-microgrid framework combining carbon capture and utilization
  technologies with P2P energy trading, trained using the MAPPO algorithm to
  reduce both cost and carbon emissions. In addition, a recurrent neural network (RNN)-based MAPPO
  algorithm was presented in \cite{zhouJointEnergyCarbon2024} for learning one-to-one
  clearing policies in the coupled energy and carbon trading markets. 

  In general, existing studies on multi-microgrid energy trading either rely on explicitly formulated optimization models or adopt learning-based approaches under simplified market and decision structures. Many of these studies assume cooperative behaviors, restricting the coupling between market bidding and energy storage decisions, or incorporate carbon considerations primarily through exogenous constraints or pricing components. Instead, this paper focuses on a decentralized intraday P2P trading setting in which autonomous microgrids make strategic trading and energy storage decisions through a market mechanism that jointly shapes incentives and trading outcomes under uncertainty.

  \section{Problem Statement}
\begin{figure}
    \centering
    \includegraphics[height=0.5\textwidth, width=0.48\textwidth]{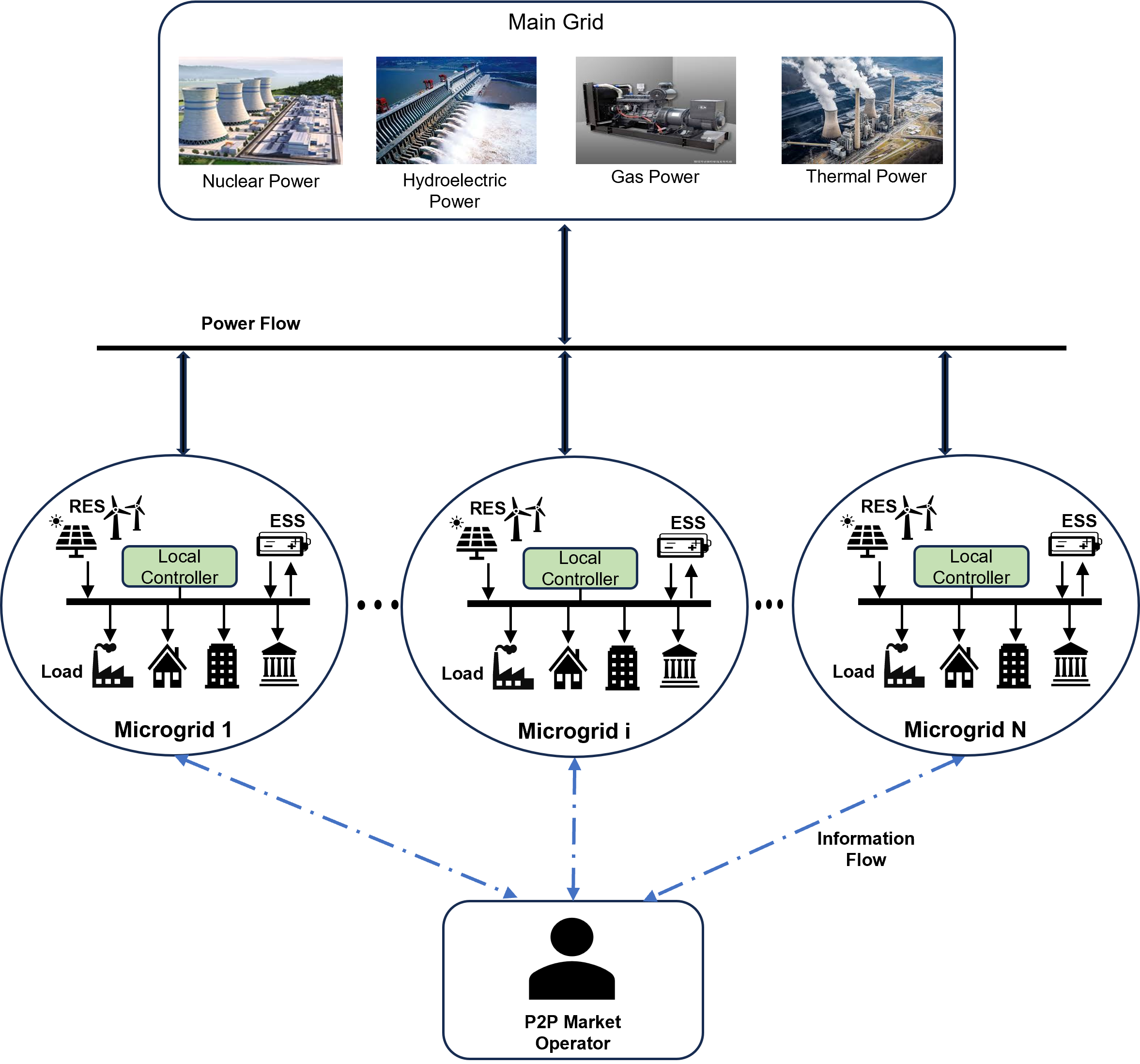}
    \caption{Structure of the power distribution network.}
    \label{fig:market_framework}
  \end{figure}

  Consider a power distribution network comprising a main grid and a set $\mathcal{N}
  = \{1, \dots, N\}$ of interconnected microgrids, as illustrated in Fig.~\ref{fig:market_framework}.
  These microgrids represent diverse prosumer archetypes, including residential, industrial, and commercial sectors, characterized by distinct load profiles and energy behaviors. The main grid functions as an external energy supplier,
  aggregating heterogeneous generation resources that range from thermal units to large-scale clean energy units. Consequently, electricity procurement from the
  main grid is characterized by time-varying intensities of carbon emission,
  necessitating a trade-off between community economic welfare (profit maximization)
  and societal welfare (carbon footprint reduction).

  Each microgrid $i \in \mathcal{N}$ operates as an autonomous agent equipped
  with inflexible electrical loads, local Distributed Generation (DG) units such
  as Photovoltaics (PV), and ESS.
  These assets provide the operational flexibility required for arbitrage and
  resilience against supply uncertainties. To maintain tractability and prioritize
  the analysis of the proposed P2P market mechanism and carbon accounting framework,
  the validation of physical power-flow constraints (e.g., line flows and AC power-flow equations) is omitted. In this specific study, we focus on active power balance and economic transactions, while transmission line congestion and reactive power flows
  are not considered.

  \subsection{Microgrid System}
  For each microgrid $i$, let $L_{i,t}$ denote the inelastic electricity demand
  and $G_{i,t}$ denote the generation of renewable power (e.g. PV output) at time
  slot $t$. The ESS serves as
  a critical flexibility asset, allowing the microgrid to arbitrage energy
  prices and maintain power balance under uncertainties. Let $E_{i,t}$ denote the stored
  energy level (kWh) and $T^{\text{ESS}}_{i,t}$ denote the charging or discharging
  rate (kW) at time $t$. A positive $T^{\text{ESS}}_{i,t}> 0$ indicates
  charging, while a negative value indicates discharging. The state-of-charge (SoC)
  evolution is governed by:
  \begin{equation}
    \label{eq: SoC}E_{i, t+1}=
    \begin{cases}
      E_{i,t}+ \dfrac{\eta_{\mathrm{ch}} T^{\text{ESS}}_{i,t} \Delta t}{E_{i, \max}},  & \text{if }T^{\text{ES}}_{i,t}\geq 0, \\
      E_{i,t}+ \dfrac{T^{\text{ESS}}_{i,t} \Delta t}{\eta_{\mathrm{dis}} E_{i, \max}}, & \text{if }T^{ES}_{i,t}< 0,
    \end{cases}
  \end{equation}
  where $\Delta t$ represents the time interval. The parameters
  $\eta_{\mathrm{ch}}$ and $\eta_{\mathrm{dis}}$ represent the charging and
  discharging rates, respectively. To ensure safe operation, the ESS is subject
  to the following constraints:
  \begin{equation}
    \label{eq: ESconstraints}
    \begin{aligned}
      \underline{T}^{\text{ESS}}_{i} & \le T^{\text{ESS}}_{i,t}\le \overline{T}^{\text{ESS}}_{i,t}, \\
      E_{i, \min}                    & \le E_{i, t}\le \alpha_{i, t}^{E}E_{i, \max},
    \end{aligned}
  \end{equation}
  where $\underline{T}^{\text{ESS}}_{i}$ and $\overline{T}^{\text{ESS}}_{i}$
  denote the maximum discharging (negative) or charging (positive) rates,
  and $E_{i, \min}$ and $E_{i, \max}$ denote the minimum and maximum energy capacity, respectively. $\alpha_{i, t}^{E}\in [0,1]$ is a control parameter for capacity reservation.

  \subsection{Electricity Market Framework}
  \label{subsection: eletricitymarket} In this work, we employ a two-stage electricity
  settlement framework comprising a Day-Ahead (DA) stage and an Intraday real-time
  stage. The DA stage functions as the primary scheduling phase, where
  microgrids secure base-load electricity requirements based on forecasts of
  demands and renewable generation. Subsequently, the Intraday P2P market serves
  as a balancing mechanism, allowing agents to trade deviations caused by uncertainties, thereby reducing imbalance penalties.

  \subsubsection{Day-Ahead Market Mechanism}
  In the DA stage, the main grid publishes three distinct price signals: the
  standard day-ahead buying price $p^{\text{G}}_{\text{da}}$, the emergency (real-time)
  buying price $p^{\text{G}}_{\text{e}}$, and the feed-in tariff (FiT)
  $p^{\text{G}}_{\text{f}}$. To reflect the marginal cost of regulation and carbon
  intensity, the pricing structure follows the inequality:
  \begin{equation}
    \label{eq:prices}p^{\text{G}}_{\text{f}}\le p^{\text{G}}_{\text{da}}\le p^{\text{G}}
    _{\text{e}}.
  \end{equation}
  This hierarchy is designed to penalize reliance on real-time emergency power (often supplied by carbon-intensive peaker plants) and to discourage excessive reverse power flow. In this work, we assume $p^{\text{G}}_{\text{f}}$ is constant, while $p^{\text{G}}_{\text{e}}$ varies dynamically to reflect the scenario of real-time grid in reality.

  Each microgrid $i$ estimates its hourly load $\bar{L}_{i,t}$ and renewable generation $\bar{G}_{i,t}$ for the upcoming operating day. Due to limited generation capacity and unavoidable forecast errors, microgrids implement a day-ahead procurement policy to cover expected deficits. The quantity of electricity purchased from the
  main grid in the DA stage, $q^{\mathrm{da}}_{i,t}$, is determined by:
  \begin{equation}
    \label{eq:dayaheadpolicy}q^{\mathrm{da}}_{i,t}= \max \left\{0, \beta_{i}(\bar
    {L}_{i,t}- \bar{G}_{i,t}) \right\},
  \end{equation}
  where $\beta_{i}> 0$ represents the procurement preference factor (or safety margin) for the microgrid $i$. The higher the value of $\beta_{i}$, the more conservative
  the strategy. Note that in this proposed framework, active selling to the main grid is restricted to the real-time stage, while the DA stage focuses solely on
  securing expected deficit coverage. Since $\beta_{i}$ is treated as a fixed characteristic of the microgrid, the focus of this study lies in the
  \textit{Intraday P2P trading strategy} to manage the residual energy
  imbalances.

  \subsection{Intraday P2P Market Mechanism}
  To mitigate real-time power imbalances caused by forecast uncertainties, we establish
  a community P2P market governed by a \textbf{Double Auction} clearing mechanism. While the trading decision-making is decentralized, the market clearing follows a hybrid architecture with centralized matching to improve
  the matching efficiency, consistent with established P2P frameworks \cite{haggiMultiRoundDoubleAuctionEnabled2021,
  qiuMultiagentReinforcementLearning2021, yinMultiAgentDeepReinforcement2025}.

  At each time step $t$, the P2P trading process follows a sequential three-phase
  protocol: \textbf{Quotation Submission}, \textbf{Market Clearing}, and \textbf{Settlement
  and Balancing}.

  \subsubsection{Quotation Submission}
  In this framework, the trading role is encoded in the sign of the bidding price
  $p_{i,t}$. A non-negative value ($p_{i,t}\geq 0$) indicates the microgrid acts
  as a \textbf{buyer}, while a negative value ($p_{i,t}< 0$) indicates that it acts as a \textbf{seller}. The bidding quantity $q_{i,t}\ge 0$ represents the magnitude of the energy to be traded. To ensure economic rationality, the absolute
  bidding price must fall within the main grid's price envelope:
  \begin{equation}
    \label{eq:price_bounds}p^{\text{G}}_{\text{f}}\le |p_{i,t}| \le p^{\text{G}}_{\text{e}}
    .
  \end{equation}
  This constraint ensures that P2P trading remains mutually beneficial compared to
  interacting with the main grid. The bidding quantity is bounded by
  \begin{equation}
    \label{eq:bidding_amount}0 \le q_{i,t}\le \bar{Q}_{i,t},
  \end{equation}
  where $\bar{Q}_{i,t}$ represents the physical limit based on the agent's role:
  \begin{equation}
    \label{eq: maxbidddingamount}\bar{Q}_{i,t}=
    \begin{cases}
      \max \{0, L_{i, t}- G_{i, t}+ \overline{T}^{\text{ESS}}_{i}\},  & \text{if }p_{i,t}\geq 0, \\
      \max \{0, G_{i, t}- L_{i, t}- \underline{T}^{\text{ESS}}_{i}\}, & \text{if }p_{i,t}< 0.
    \end{cases}
  \end{equation}
  This formulation allows microgrids to flexibly switch roles and optimize their
  quotations to maximize individual utility.

  \subsubsection{Market Clearing}
  The operator aggregates all buy and sell orders, executes the matching
  algorithm, and determines the clearing price and traded quantities. To
  effectively balance the real-time demand/supply of the multi-microgrid systems, a new
  bidding matching mechanism is proposed with the \textit{mid-point pricing rule}.
  This rule guarantees \textit{strict budget balance}, ensuring that the market operator acts purely as a neutral facilitator without retaining any surplus. Such
  neutrality is essential for maintaining trust and sustaining long-term
  participation in P2P energy trading.

  \begin{algorithm}
    [t]
    \caption{Joint Price--Quantity (JPQ) Market Clearing Mechanism}
    \label{alg:marketclearing}
    \begin{algorithmic}
      [1] \REQUIRE Quotation set $\mathcal{Q}$, System Imbalance $m_{t}$, Emergency
      Price $p_{e}^{\text{G}}$ \ENSURE Trading Quantity Matrix $\mathbf{Q}$,
      Trading Price Matrix $\mathbf{\Pi}$ \STATE \textbf{Initialization:} \STATE
      Partition agents into Buyers $\mathcal{B}$ and Sellers $\mathcal{S}$ \STATE
      Sort $\mathcal{B}$ and $\mathcal{S}$ using Algorithm~\ref{alg:quotation-sorting}
      \STATE Initialize matrices $\mathbf{Q}\leftarrow \mathbf{0}$,
      $\mathbf{\Pi}\leftarrow \mathbf{0}$ \STATE Initialize pointers: $b, s, b_{\text{start}}
      , s_{\text{start}}\leftarrow 1$
      \vspace{0.5ex}
      \STATE \textbf{Matching Process:} \WHILE{$b \le |\mathcal{B}|$ \AND $s \le |\mathcal{S}|$}

      \IF{$p_{b,t}< |p_{s,t}|$} \IF{$m_{t}< 0$} \STATE
      $b \leftarrow b + 1, \quad b_{\text{start}}\leftarrow b_{\text{start}}+ 1$
      \STATE \textbf{continue} \ELSIF{$m_{t}> 0$} \STATE
      $s \leftarrow s + 1, \quad s_{\text{start}}\leftarrow s_{\text{start}}+ 1$
      \STATE \textbf{continue} \ELSE \STATE \textbf{break} \ENDIF \ENDIF

      \STATE $\mathbf{Q}_{b,s}\leftarrow \min(q_{b,t}, q_{s,t})$,
      $\mathbf{\Pi}_{b,s}\leftarrow (p_{b,t}+ |p_{s,t}|)/2$

      \STATE $q_{b,t}\leftarrow q_{b,t}- \mathbf{Q}_{b,s}, \quad q_{s,t}\leftarrow
      q_{s,t}- \mathbf{Q}_{b,s}$

      \IF{$q_{b,t}= 0$} \STATE $b_{\text{start}}\leftarrow b_{\text{start}}+ 1$ \ENDIF
      \IF{$q_{s,t}= 0$} \STATE $s_{\text{start}}\leftarrow s_{\text{start}}+ 1$
      \ENDIF

      \STATE $b \leftarrow b + 1, \quad s \leftarrow s + 1$ \IF{$b > |\mathcal{B}|$}
      \STATE $b \leftarrow b_{\text{start}}$ \ENDIF \IF{$s > |\mathcal{S}|$} \STATE
      $s \leftarrow s_{\text{start}}$ \ENDIF \ENDWHILE

      \RETURN $\mathbf{Q}$, $\mathbf{\Pi}$
    \end{algorithmic}
  \end{algorithm}

  Building on the double-auction foundations in \cite{haggiMultiRoundDoubleAuctionEnabled2021},
  \textbf{Algorithm~\ref{alg:marketclearing}}
  is designed to produce economically feasible and system-aligned trades among
  microgrids. The clearing process adheres to three main design principles:
  \begin{itemize}
    \item \textbf{Individual Rationality} -- A trade is executed only when the buyer’s
      bid price is no less than the seller’s ask ($p_{b,t}\ge |p_{s,t}|$),
      ensuring voluntary participation and preventing economically inefficient
      exchanges.

    \item \textbf{System Incentives} -- In sorting algorithm (\textbf{Algorithm~\ref{alg:quotation-sorting}}),
      the mechanism incorporates the real-time market factor $m_{t}$ to prioritize
      trades that support system-level needs.
      \begin{itemize}
        \item Under a \textit{surplus} ($m_{t}< 0$), the mechanism favors buyers
          with greater absorption capability rather than simply selecting the highest
          bidders. This encourages microgrids to absorb excess renewable
          generation.

        \item Under a \textit{deficit} ($m_{t}> 0$), the mechanism prioritizes sellers
          capable of providing the largest welfare contribution, incentivizing
          energy sharing to enhance community stability.
      \end{itemize}

    \item \textbf{Equitable Matching} -- By iteratively examining feasible buyer-seller
      pairs, the mechanism ensures that microgrids submitting valid quotations
      receive fair access to trading opportunities. This prevents local matching
      bias and promotes a more balanced allocation of liquidity across participants.
  \end{itemize}

  \begin{algorithm}
    [t]
    \caption{Market-Driven Quotation Sorting}
    \label{alg:quotation-sorting}
    \begin{algorithmic}
      [1] \REQUIRE Sets of buyers $\mathcal{B}$ and sellers $\mathcal{S}$,
      Market factor $m_{t}$, Emergency Price $p_{e}^{\text{G}}$ \ENSURE Sorted sets
      $\mathcal{B}$ and $\mathcal{S}$
      \vspace{0.5ex}
      \STATE \textbf{Initialize Sorting Metrics:} \STATE For each buyer
      $i \in \mathcal{B}$: $\kappa_{1}^{B}\leftarrow p_{i,t}$, \quad
      $\kappa_{2}^{B}\leftarrow p_{i,t}\cdot q_{i,t}$ \STATE For each seller
      $j \in \mathcal{S}$: $\kappa_{1}^{S}\leftarrow |p_{j,t}|$, \quad
      $\kappa_{2}^{S}\leftarrow (p_{e}^{\text{G}}- |p_{j,t}|) \cdot q_{j,t}$
      \vspace{0.5ex}
      \STATE \textbf{Execute Sorting based on market factor $m_{t}$:} \IF{$m_{t}< 0$ (Electricity Surplus)}
      \STATE Sort $\mathcal{B}$ descending by $\kappa_{2}^{B}$ \STATE Sort
      $\mathcal{S}$ ascending by $\kappa_{1}^{S}$ \ELSIF{$m_{t}> 0$ (Electricity Deficit)}
      \STATE Sort $\mathcal{B}$ descending by $\kappa_{1}^{B}$ \STATE Sort $\mathcal{S}$
      descending by $\kappa_{2}^{S}$ \ELSE \STATE Sort $\mathcal{B}$ descending by
      $\kappa_{1}^{B}$ \STATE Sort $\mathcal{S}$ ascending by $\kappa_{1}^{S}$ \ENDIF
      \RETURN Ordered sets $\mathcal{B}$ and $\mathcal{S}$
    \end{algorithmic}
  \end{algorithm}

  \subsubsection{Settlement and Balancing}
  After market clearing, the settlement phase executes both physical delivery and financial compensation. Buyers compensate sellers according to the clearing price matrix $\mathbf{\Pi}$, and sellers inject the commitment power quantities into buyers based on $\mathbf{Q}$.

  Because P2P trades may not fully eliminate real-time imbalances (due to market frictions, renewable energy uncertainties, or physical constraints), each microgrid
  adopts a hierarchical recourse strategy to resolve any residual deviations:
  \begin{enumerate}
    \item \textit{Self-Correction:} The microgrid first uses its local ESS to supply the deficit, subject to the SoC and related constraints in \eqref{eq: ESconstraints}.

    \item \textit{Emergency Procurement:} If a deficit persists after ESS discharge, the microgrid purchases emergency energy $q^{\text{e}}_{i,t}$ from the main grid at the penalty price $p_{e}^{\text{G}}$.

    \item \textit{Feed In Strategy:} For microgrid $i$, the total export quantity $q^{\text{fit}}
      _{i,t}$ is compensated at the feed-in tariff $p^{\text{G}}_{f}$, where
      $q^{\text{fit}}_{i,t}$ includes both the energy discharged to reduce over-storage
      and the portion of renewable surplus that cannot be accommodated by the
      ESS.
  \end{enumerate}

  Under this recourse structure, the real-time power balance for each microgrid $i$
  at time $t$ satisfies:
  \begin{equation}
    \label{eq:output_input}L_{i,t}+ q^{\text{fit}}_{i,t}+ q^{\text{s}}_{i,t}+ T_{i,t}
    ^{\text{ESS}}= G_{i,t}+ q^{\text{da}}_{i,t}+ q^{\text{b}}_{i,t}+ q^{\text{e}}
    _{i,t},
  \end{equation}
  where $q^{\text{b}}_{i,t}= \sum_{j}\mathbf{Q}_{(}i,j)$ and
  $q^{\text{s}}_{i,t}= \sum_{j}\mathbf{Q}(j,i)$ denote the microgrid’s total P2P
  purchase and sale quantities derived from $\mathbf{Q}$. The terms $P^{\text{ch}}
  _{i,t}$ and $P^{\text{dis}}_{i,t}$ represent the charging and discharging power
  of the local ESS, respectively.

  \section{MARL-based Bidding Strategy in Intraday P2P Market}
  In this section, the Intraday P2P trading model mentioned above can be
  formulated as a DEC-POMDP, which is typically solved using MARL algorithms. In
  this work, we develop an \textbf{LSTM-MAPPO} framework to handle the high-dimensional,
  sequential, and interactive nature of this decision-making task. The proposed architecture
  follows the CTDE paradigm. Specifically, during training, a centralized critic leverages global
  information to provide stable value estimation and reduce gradient variance,
  thereby improving learning efficiency in a competitive multi-agent environment.
  During execution, however, each microgrid behaves autonomously, generating its
  bidding and storage decisions using only its own local observation history. This
  ensures both data privacy and scalability, while enabling robust policy
  learning in the stochastic and strategically complex P2P market.

  \subsection{Bidding and Energy Storage Strategies for Microgrid $i$}
  In the Intraday P2P market, each microgrid $i$ seeks to maximize its own economic
  benefit by optimally selecting its bidding price, bidding quantity, and ESS reservation
  level. The profit-maximization problem of microgrid $i$ is formulated as:
  \begin{align}
    \max_{p_{i,t}, q_{i,t}, \alpha_{i,t}^E}\quad & \sum_{t=1}^{T}\left( P^{\text{G}}_{i,t}+ P^{\text{p2p}}_{i,t}\right) \label{eq:obj}                  \\
    \text{s.t.}\quad                              \\
                                                 & p^{\text{G}}{\text{f}}\le p_{i,t}\le p^{\text{G}}{\text{e}}, \label{eq:price}                       \\
                                                 & 0 \le \alpha_{i,t}^{E}\le 1, \label{eq:alpha}                                                       \\
                                                 & \text{Constraints }~\eqref{eq: ESconstraints},~\eqref{eq:bidding_amount},~\eqref{eq:output_input}.
  \end{align}
  Here, $P^{\text{G}}_{i,t}$ captures the net profit of microgrid $i$ from
  interacting with the main grid, while $P^{\text{p2p}}_{i,t}$ represents its profit
  from P2P transactions. Because the expected day-ahead purchase cost is
  constant when $\beta_{i}$ is fixed, the main-grid profit $P^{\text{G}}_{i,t}$ only
  depends on feed-in and emergency operations:
  \begin{equation}
    P^{\text{G}}_{i,t}= p^{\text{G}}_{\text{f}}q^{\text{fit}}_{i,t}- p^{\text{G}}
    _{\text{e}}q^{\text{e}}_{i,t}.
  \end{equation}
  Similarly, the P2P profit of microgrid $i$ is expressed as
  \begin{equation}
    P^{\text{p2p}}_{i,t}= \sum_{j \in n_b}p_{i,j,t}q_{i,j,t}- \sum_{j \in n_s}p_{i,j,t}
    q_{i,j,t},
  \end{equation}
  where $n_{b}$ and $n_{s}$ denote the sets of matched buyers and sellers for microgrid
  $i$, with $n_{b}+ n_{s}\le N$.

  Since the P2P clearing mechanism is strictly budget-balanced, the total P2P profit
  across all microgrids is zero. Hence, the aggregate economic welfare of the community
  depends solely on the main-grid interactions. The system-level objective is
  therefore
  \begin{equation}
    \label{eq:total_benefit}\max \sum_{t=1}^{T}\sum_{i \in \mathcal{N}}P^{\text{G}}
    _{i,t}.
  \end{equation}
  The microgrids’ profit-maximizing bidding interactions form a decentralized
  and partially observable multi-agent decision process, motivating the subsequent
  formulation of the Intraday market as a DEC-POMDP.

  \subsection{DEC-POMDP Formulation}

  The Intraday P2P trading problem can be modeled as a DEC-POMDP. A DEC-POMDP is
  defined by the tuple
  \[
    \left( \mathcal{N},\, \mathcal{S},\, \mathcal{A},\, \mathcal{O},\, \mathcal{R}
    ,\, \mathcal{P},\, \gamma \right),
  \]
  where the components are described as follows.

  \subsubsection{Agents}
  Each agent $i \in \mathcal{N}$ corresponds to a microgrid participating in the
  P2P energy market. Agents act simultaneously and interact through the market-clearing
  mechanism and the physical power flow balance.

  \subsubsection{Global State}
  The global environment state at time $t$ is denoted by
  \[
    \mathcal{S}_{t}= \{ s_{1,t},\, s_{2,t},\, \dots,\, s_{N,t}\},
  \]
  where $s_{i,t}$ includes the internal operational variables of microgrid $i$
  and aggregated market conditions. The global state is not directly observable to
  any individual agent.

  \subsubsection{Observation}
  At each time step, the observation $\mathcal{O}_{t}$ denotes the set of all
  agents’ observations. For each agent $i$, it receives a local observation
  \[
    \mathcal{O}^{i}_{t}= \{ m_{t}, E_{i,t}, h_{i,t}, t \},
  \]
  where $m_{t}$ is the market factor summarizing the supply–demand condition, $E_{i,t}$
  is the SOC of microgrid $i$, and
  \[
    h_{i,t}= \{ q^{\mathrm{da}}_{i,z},\, \bar{L}_{i,z},\, \bar{G}_{i,z},\, p^{G}_{e,z}
    \}_{z=t-\delta_1}^{t+\delta_2}
  \]
  is a noisy temporal observation window containing historical and forecasted
  information on day-ahead purchases, load, renewable generation, and emergency prices.
  This window allows agents to incorporate temporal correlations and prediction
  uncertainty into decision-making.

  \subsubsection{Actions}
  At time $t$, the joint action of all agents is
  \[
    \mathcal{A}_{t}= \{ \mathcal{A}^{1}_{t}, \dots, \mathcal{A}^{N}_{t}\}.
  \]
  For each agent $i$, it selects the action
  \[
    \mathcal{A}^{i}_{t}= \{ p_{i,t},\, q_{i,t},\, \alpha_{i,t}^{E}\},
  \]
  where $p_{i,t}$ and $q_{i,t}$ denote the bidding price and quantity in the
  Intraday P2P market, and $\alpha_{i,t}^{E}$ is the ESS reservation (or
  emergency procurement) decision.

  \subsubsection{Reward Function}
  After executing the joint action $\mathbf{A}_{t}$, agent $i$ receives the
  immediate reward
  \[
    \mathcal{R}^{i}_{t}= P^{\mathrm{G}}_{i,t}+ P^{\mathrm{p2p}}_{i,t},
  \]
  which reflects the microgrid’s operational economic benefit from P2P trading
  and interactions with the main grid. Similarly, let $\mathcal{R}_{t}$ denote
  rewards of all agents. The reward promotes policies that efficiently utilize renewables,
  reduce emergency purchases, and coordinate energy exchange through the P2P market.

  \subsubsection{State Transition Function}
  The environment evolves according to the transition function
  \[
    \mathcal{P}(\mathcal{S}_{t+1}\mid \mathcal{S}_{t},\, \mathbf{A}_{t}),
  \]
  which jointly captures (i) the clearing results of the P2P market, (ii) the
  physical power balance and ESS dynamics in \eqref{eq:output_input}, and (iii)
  exogenous uncertainties such as renewable variability and load fluctuations. This
  transition rules how local actions collectively shape the global system evolution.

  \subsubsection{Discount Factor}
  The discount factor $\gamma \in (0,1)$ determines each agent’s preference over
  future rewards.

  \subsection{LSTM-MAPPO with CTDE Training Method}
  In this work, the LSTM-MAPPO algorithm is presented to address P2P trading
  problem among $N$ microgrids. The LSTM-MAPPO framework consists of $N$ decentralized
  policy networks, denoted by $\{\pi_{\theta_i}\}_{i=1}^{N}$, and critic networks
  $\{V_{\phi_i}\}_{i=1}^{N}$, which are used to approximate the policy and value
  functions of the microgrids, respectively.

\begin{figure}
    \centering
    \includegraphics[width=0.48\textwidth]{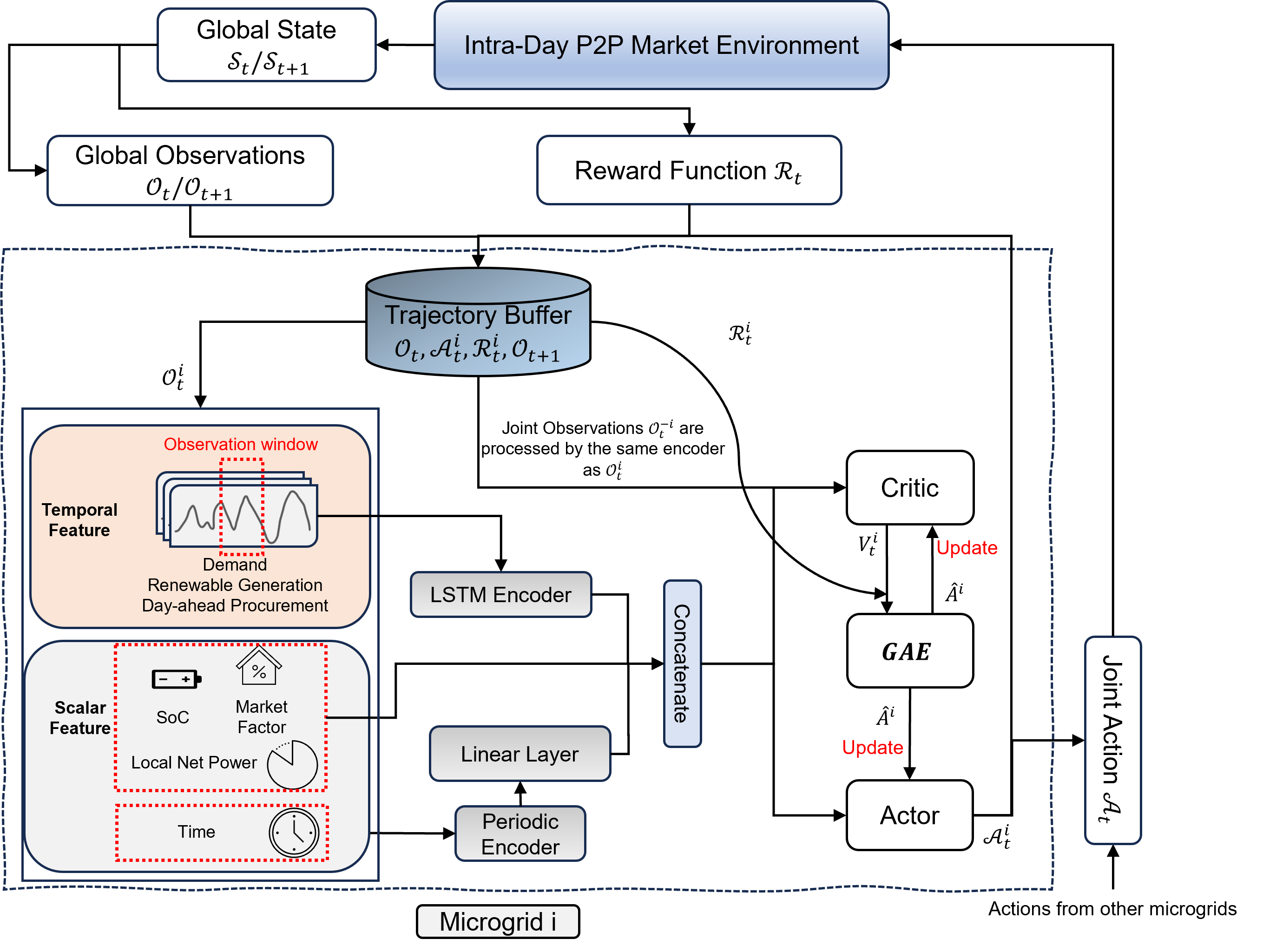}
    \caption{The training process of LSTM-MAPPO algorithm for Intraday P2P
    market.}
    \label{fig:training_model}
  \end{figure}

  Each agent $i$ selects the action $\mathcal{A}^{i}_{t}$ based on local observation
  $\mathcal{O}^{i}_{t}$ and the policy $\pi_{\theta_i}$ to maximize the discounted
  cumulative reward $J^{i}(\theta_{i})$ and learn a value function $V_{\phi_i}$ based
  on the global observation $\mathcal{O}_{t}$. The policy $\pi_{\theta_i}$ of
  agent $i$ is updated by minimizing the following loss function:
  \begin{equation}
    \label{eq:policy_update}
    \begin{aligned}
      \mathcal{L}_{\text{actor}}(\theta_{i}) = -\,\frac{1}{B}\sum_{l=1}^{B}\Big[ & \min\!\big( \rho_{\theta_i, l}A^{i}_{l},\; \mathrm{clip}(\rho_{\theta_i, l},\, 1-\epsilon,\, 1+\epsilon) \\
      & A^{i}_{l}\big) \;-\; c \mathcal{H}\!\left(\pi_{\theta_i}(\cdot \mid \mathcal{O}^{i}_{l})\right)\Big],
    \end{aligned}
  \end{equation}
  where $B$ is the batch size, $c$ is the
  entropy parameter, $\epsilon$ is the clipping threshold, and
  $\mathcal{H}[\cdot]$ is the entropy of the policy distribution, and $\rho_{\theta_i, l}$ is the importance sampling ratio
  \begin{equation}
    \label{eq:policy_ratio}\rho_{\theta_i, l}= \frac{\pi_{\theta_i}\big(\mathcal{A}^{i}_{t}|
    \mathcal{O}^{i}_{t}\big)}{\pi_{\theta_i^{\text{old}}}\big(\mathcal{A}^{i}_{t}|
    \mathcal{O}^{i}_{t}\big)},
  \end{equation}
  where $\pi_{\theta_i^{\text{old}}}$ denotes the old policy of agent $i$ used to generate the samples in the current batch. And the advantage
  $A^{i}_{l}$ is computed by \textit{Generalized Advantage Estimation} (GAE):
  \begin{equation}
    \label{eq:gae_def}\hat{A}^{i}_{l}= \sum_{t=0}^{T-1}(\gamma \lambda)^{t}\left
    ( \mathcal{R}^{i}_{l+t}+ \gamma V_{\phi_i}(s_{i, l+t+1}) - V_{\phi_i}(s_{i, l+t}) \right),
  \end{equation}
  where $\lambda$ denotes the GAE parameter. The critic is updated by minimizing the regression error between the predicted
  value and the GAE–based target. The critic loss for agent $i$ is defined as
  \begin{equation}
    \label{eq:critic_loss}\mathcal{L}_{\text{critic}}(\phi_{i}) = \frac{1}{B}\sum
    _{l=1}^{B}\left( V_{\phi_i}(\mathcal{O}^{i}_{l}) \;-\; \hat{V}^{i}_{l}\right
    )^{2},
  \end{equation}
  where $\hat{V}^{i}_{l}= A^{i}_{l}+ V_{\phi_i}^{\text{old}}(\mathcal{O}^{i}_{l})$ is the target constructed
  from the GAE and the baseline value while $V_{\phi_i}^{\text{old}}$ denotes the old critic network of agent $i$ when the samples were collected. Given the learning rates $\alpha_{\theta}$
  and $\alpha_{\phi}$, the parameters of policy and critic networks of agent $i$
  are updated based on gradient information:
  \begin{equation}
    \begin{aligned}
       & \theta_{i}\leftarrow \theta_{i}- \alpha_{\theta}\,\nabla_{\theta_i}\mathcal{L}_{\text{actor}}(\theta_{i}), \\
       & \phi_{i}\leftarrow \phi_{i}- \alpha_{\phi}\,\nabla_{\phi_i}\mathcal{L}_{\text{critic}}(\phi_{i}).
    \end{aligned}
  \end{equation}
  To enhance temporal representation, an LSTM module is incorporated for
  sequential observation embedding, while a sinusoidal periodic encoder is employed
  to capture daily cyclical patterns. The overall training process of MMAPPO is
  summarized in Algorithm~\ref{alg:LSTM-Mappo}. For further architectural and implementation
  details of the policy and critic networks, readers are referred to
  \cite{yu_surprising_2022}.
  \begin{algorithm}
    \caption{LSTM-MAPPO for P2P Trading}
    \label{alg:LSTM-Mappo}
    \begin{algorithmic}
      [1] \STATE Initialize simulation parameters. \STATE Initialize policy
      networks $\{\pi_{\theta_i}\}_{i=1}^{N}$, critic networks $\{V_{\phi_i}\}_{i=1}
      ^{N}$, and the set of buffers $\mathcal{D}= \{\mathcal{D}_{i}\}_{i=1}^{N}$.
      \FOR{$episode=1$ to $N_{\mathrm{ep}}$} \STATE Reset the environment and
      Clear all buffers. \FOR{$t=1$ to $T$} \STATE Actions sampling
      $\{\mathcal{A}^{i}_{t}\sim \pi_{\theta_i}(\cdot|\mathcal{O}^{i}_{t})\}_{i=1}
      ^{N}$. \STATE Environment steps $\to \mathcal{S}_{t+1}$, $\mathcal{R}_{t}$.
      \STATE Store $\{\mathcal{O}_{t},\mathcal{A}_{t},\mathcal{R}_{t},\mathcal{O}
      _{t+1}, \mathcal{S}_{t},\mathcal{S}_{t+1}\} \to \mathcal{D}$. \ENDFOR \FOR{$epoch=1$ to $K$}
      \FOR{each mini-batch $\mathcal{B}$ from buffers} \STATE Normalized
      advantages $\{A^{i}_{t}\}_{i=1}^{N}\gets V_{\phi_i}(s_{i,t})$. \STATE Update
      each $\theta_{i}$ via the gradient of per-agent policy loss. \STATE Update
      each $\phi_{i}$ via the gradient of per-agent critic loss. \ENDFOR \ENDFOR
      \ENDFOR
    \end{algorithmic}
  \end{algorithm}
  In training process of \textbf{Algorithm \ref{alg:LSTM-Mappo}}, it is noted that
  CTDE method is adopted to protect the privacy of each agent and improve the convergence
  performance. The training process for the multi-microgrid system is shown in Fig.~\ref{fig:training_model}.
  In addition, it requires total observations of all gents in critic training while
  only local information is needed in actor training.

  \section{Numerical Simulation}
  \subsection{Environment Setup}
  We consider an electricity market comprising four microgrids equipped with PV generation and connected to a main grid. The four microgrids serve as representative market participants—a small buyer (Grid 1), a small seller (Grid 2), a large buyer (Grid 3), and a large seller (Grid 4),—thereby capturing the diversity and heterogeneity typically observed in P2P trading environments. In addition, the bidding roles of these microgrids may shift over time due to uncertainties in demand, renewable generation, and storage conditions, reflecting the dynamic and adaptive nature of real-world P2P markets.
  The parameter settings of the four microgrids are summarized in Table~\ref{tab:grid-parameters}. In all simulations, charging and discharging processes are assumed to be lossless, i.e.,
  $\eta_{\mathrm{ch}}= \eta_{\mathrm{dis}}= 1$. The observation window for each
  microgrid is configured with $\delta_{1}= 1$ and $\delta_{2}= 6$.

\begin{figure}
    \centering
    \includegraphics[width=0.48\textwidth, height=0.4\textwidth]{
      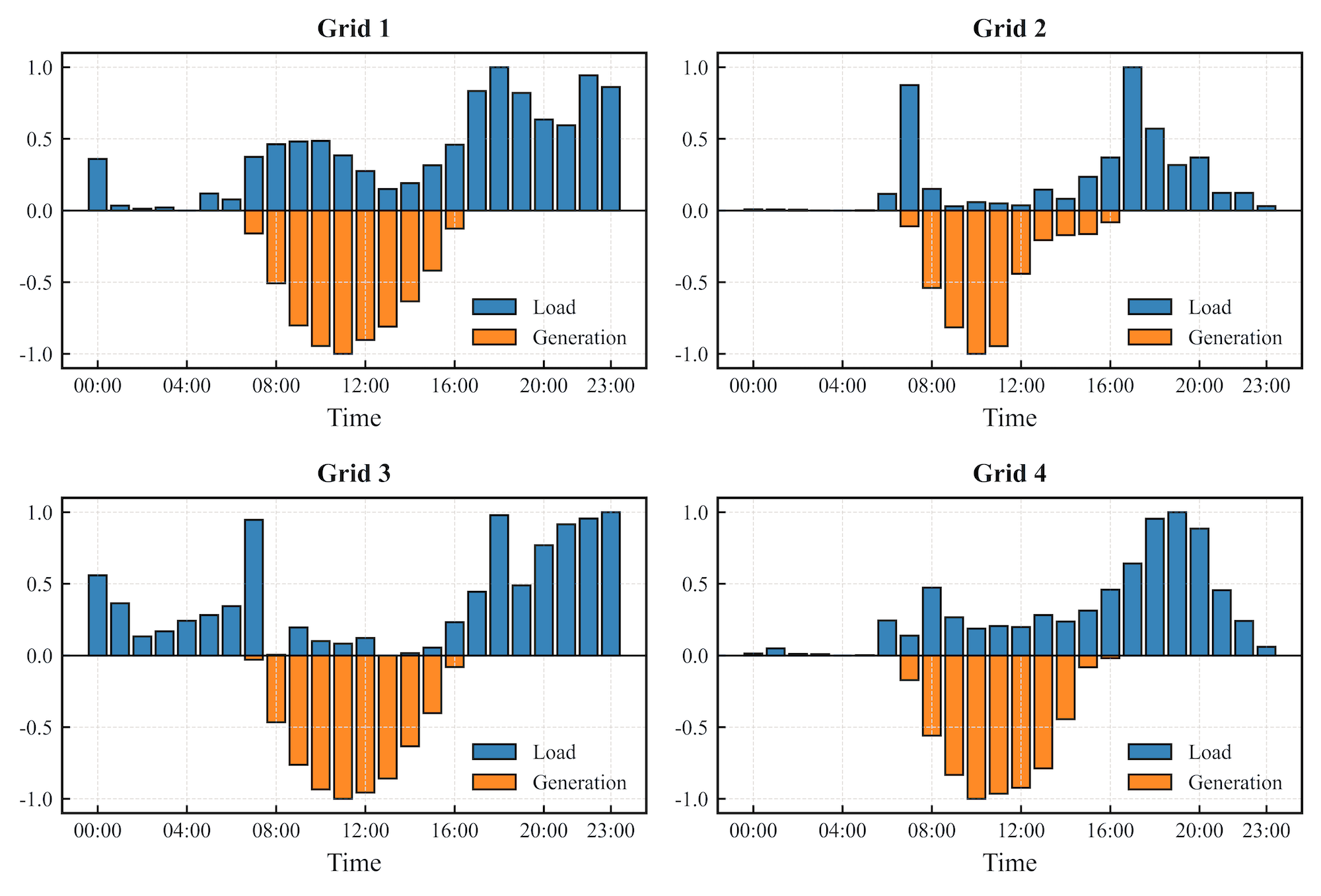
    }
    \caption{$24$-hour Normalized Demand and PV Profiles of $4$ microgrids.}
    \label{fig:demand_generation_profile}
  \end{figure}

  The daily renewable generation and electricity demand profiles of the
  microgrids are derived from the normalized dataset of four Australian
  residential households reported in \cite{ratnam2017Residential}. One year of real
  measurements is first aggregated into a representative 24-hour cycle for each household
  by averaging historical data at each hour of the day, followed by
  normalization to the interval $[0,1]$. This procedure yields realistic and numerically
  stable demand and PV generation trajectories for the constructed simulation environment.
  The resulting 24-hour demand and generation profiles for the four microgrids
  are depicted in Fig.~\ref{fig:demand_generation_profile}.

\begin{figure}
    \centering
    \includegraphics[width=0.35\textwidth]{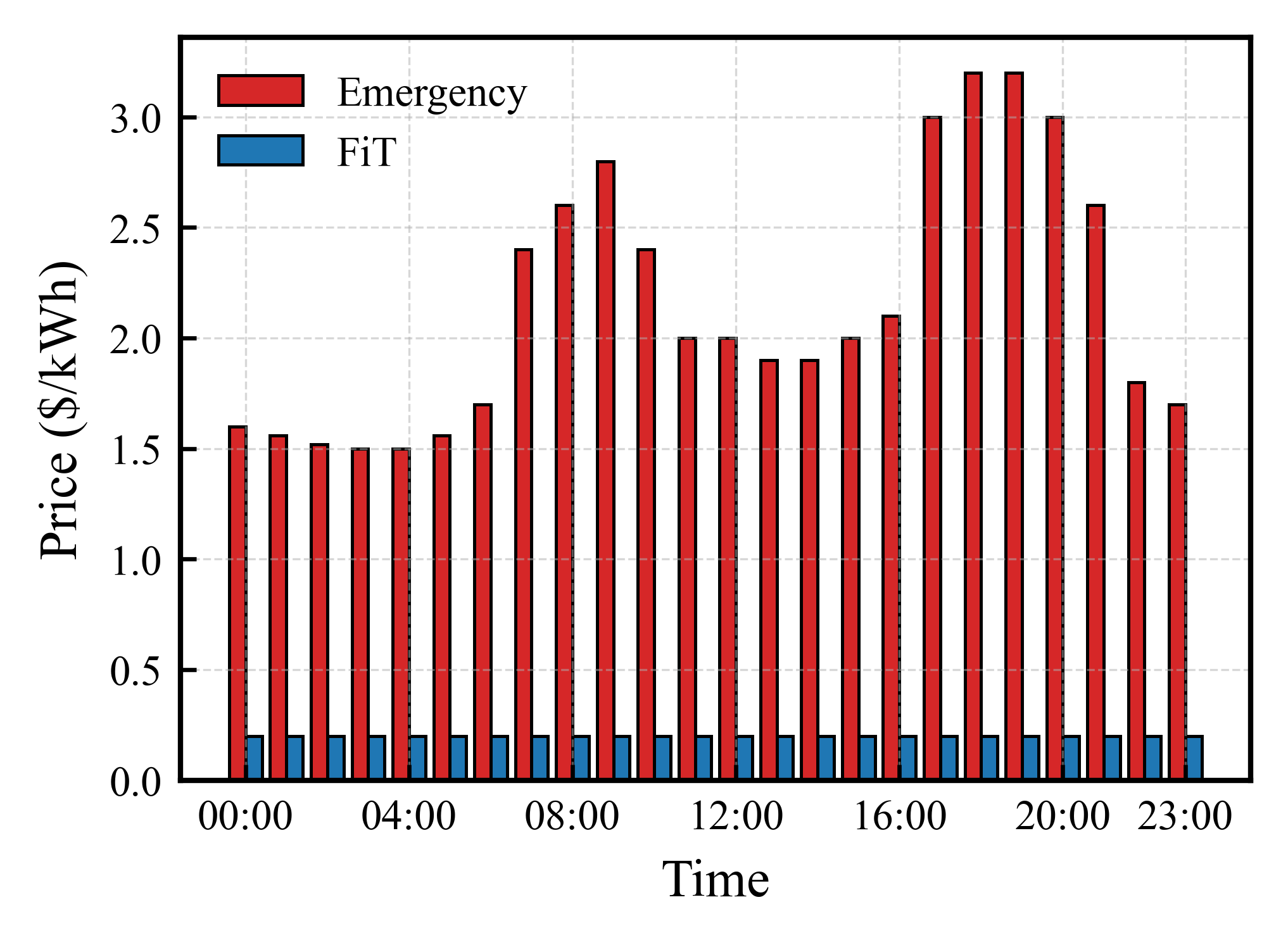}
    \caption{$24$-hour dynamical change of emergency price and FiT  over times.}
    \label{fig:dynamic_price}
  \end{figure}

  To emulate day-to-day variability observed in real-world operations, actual
  demand and generation profiles are synthesized by injecting Gaussian noise
  into the base profiles at each time step. The corresponding observations used by
  the agents are also corrupted by Gaussian noise to reflect measurement uncertainty.
  Moreover, for robustness evaluation, we model three types of PV generation
  disruptions—sudden drops, gradual declines, and complete failures—which occur
  independently with hourly probabilities of $85\%$, $10\%$, and $1\%$, respectively.
  As shown in Fig.~\ref{fig:dynamic_price}, the FiT of the main grid is
  fixed at $0.2~\$/\text{kWh}$, while the emergency price varies dynamically over
  time within the range of $[1.5, 3.5]~\$/\text{kWh}$.

  \begin{table}
    \caption{Simulation Parameters for the 4 Microgrids}
    \begin{center}
      \begin{tabular}{|c|c|c|c|c|}
        \hline
        \textbf{Parameter}                    & \textbf{Grid 1} & \textbf{Grid 2} & \textbf{Grid 3} & \textbf{Grid 4} \\
        \hline
        $L_{i}^{\text{max}}$ (kWh)            & 25              & 6               & 40              & 5               \\
        \hline
        $G_{i}^{\text{max}}$ (kWh)            & 5               & 7               & 10              & 15              \\
        \hline
        $E_{\text{max}}$ (kWh)                & 8               & 15              & 15              & 30              \\
        \hline
        $\overline{T}^{\text{ES}}_{i}$ (kW)   & 4               & 5               & 8               & 10              \\
        \hline
        $-\underline{T}^{\text{ES}}_{i}$ (kW) & 4               & 5               & 8               & 10              \\
        \hline
        $E_{0}$ (kWh)                         & 0               & 2               & 0               & 20              \\
        \hline
      \end{tabular}
      \label{tab:grid-parameters}
    \end{center}
  \end{table}

  \subsection{Baselines}
  In the simulation study, we evaluate the proposed framework from two perspectives:
  the learning algorithms and the market clearing mechanisms. 

  For the market clearing mechanisms, we compare the proposed JPQ market clearing
  mechanism with the following double auction mechanisms: (1) \textbf{Greedy} used
  in \cite{qiuMultiagentReinforcementLearning2021}; (2) \textbf{Multi-round
  double auction (MRDA)} proposed in \cite{haggiMultiRoundDoubleAuctionEnabled2021};
  (3) \textbf{Vickrey-variant double auction (VVDA)} presented in \cite{zhaoComparisonsAuctionDesigns2023}.
  
  For the learning algorithms, we compare the LSTM-MAPPO algorithm with the following MARL algorithms: (1) \textbf{MADDPG} algorithm proposed in
  \cite{qiuMultiagentReinforcementLearning2021, yinMultiAgentDeepReinforcement2025}; (2) \textbf{MADDPG-GCN} algorithm proposed in
  \cite{wengOptimizingBiddingStrategy2025}; (3) \textbf{IPPO} algorithm proposed in \cite{schulman2017proximal}; (4) \textbf{MASAC} algorithm proposed in \cite{cuiPeertopeerEnergyTrading2024,wangCollaborativeOptimizationMultimicrogrids2023}; (5) \textbf{MAPPO} algorithm proposed in \cite{yangMultistageStochasticDispatching2024,chenCombinedCarbonCapture2024}. To  fairly evaluate the performance of these algorithms, our proposed market clearing mechanism is used for the comparison of these MARL algorithms.

  \subsection{Parameter and Details of Training}
  The experimental platform uses AMD Ryzen Threadripper PRO 3955WX 16-Cores 3.9GHz
  CPU with $4$ NVIDIA GeForce RTX 3090 GPUs. The implementation of all MARL algorithms
  is based on Pytorch under Python $3.11$.

  For the LSTM-MAPPO algorithm, the trajectory buffer length is set to $T=1024$, the mini-batch size is $512$, and the number of optimization epochs is $K=10$. The actor network adopts two hidden layers of sizes $(512, 512)$, while the critic network uses hidden dimensions of $(2056, 1024)$. The LSTM encoder contains a single hidden layer with dimension $128$. The discount factor is set to $\gamma = 0.95$, and the GAE parameter is also set to $\lambda = 0.95$. ReLU is used as the activation function throughout the model.

  In the DA stage, we assume all microgrids execute the same procurement policy with $\beta_i = 0.95$ for any $i$. For the JPQ market clearing mechanism, the market factor $m_t$ is defined based on the global net power, i.e., the total demand $L_t$ minus the total PV generation $G_t$, total day-ahead procurement $q^{\text{da}}_t$ and total storage amount. Specifically, $m_t = 0$ if this index lies within the interval $[-20, -30]$, $m_t = -1$ if it is less than $-30$, and $m_t = 1$ otherwise.

  \subsection{Performance Comparison of Market Clearing Mechanism}
  \label{sim:market}
\begin{figure}
    \centering
    \includegraphics[width=0.48\textwidth]{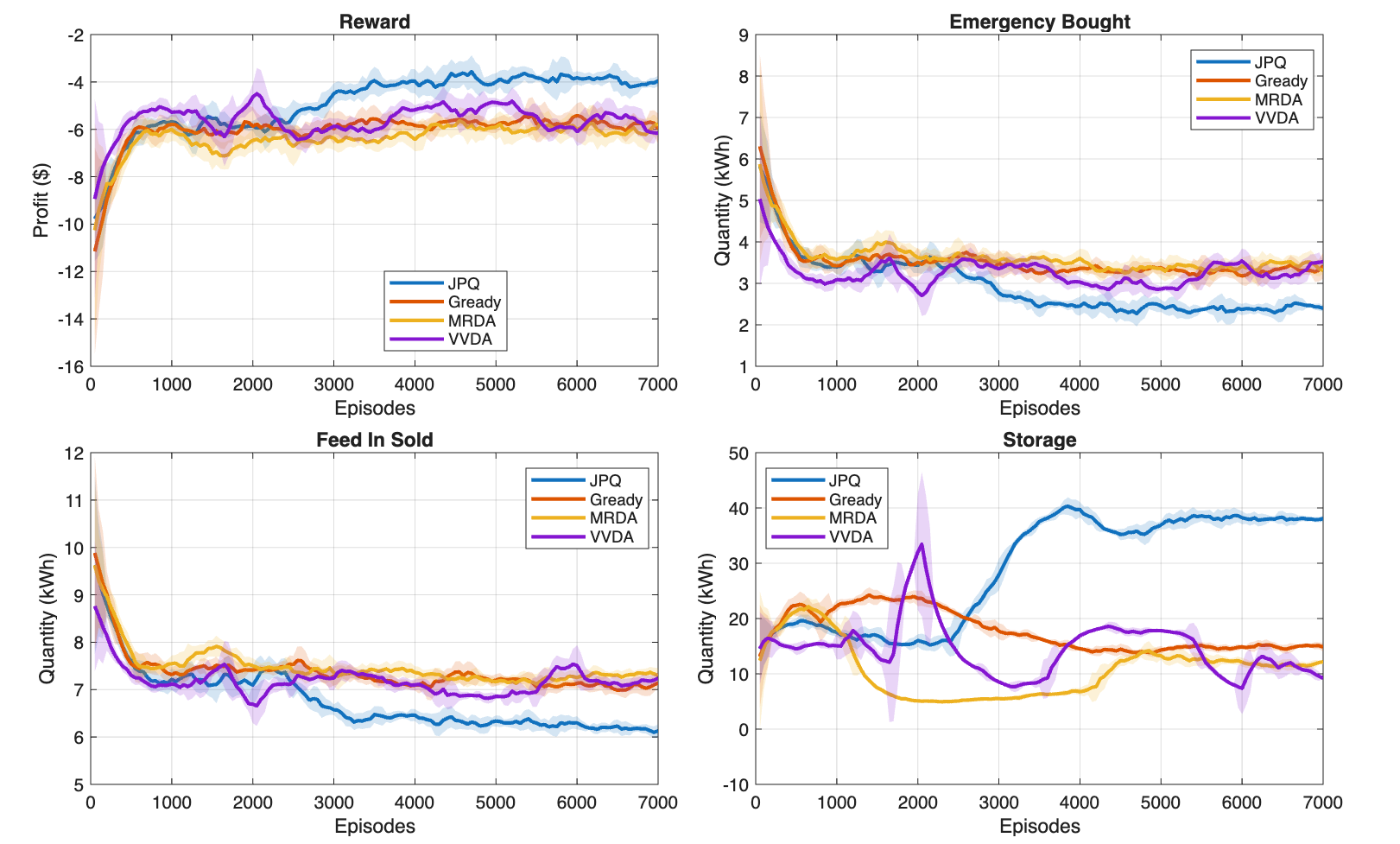}
    \caption{Training performance of four market clearing mechanism with LSTM-MAPPO algorithm over 7000 episodes. Each point denotes the hourly average across the four microgrids in a simulation trajectory.}
    \label{fig:marketcomparison}
  \end{figure}

  The performance comparison of the market clearing mechanisms for the multi-microgrid system with the LSTM-MAPPO algorithm is shown in Fig.~\ref{fig:marketcomparison} and Table.~\ref{tab:market_mechanism_comparison}. The proposed JPQ market clearing mechanism effectively promotes the community economic profit and reduce the dependencies of these microgrids on power resources with the high-carbon emission of the system. For instance, JPQ improves community profit by approximately $33\%$, reduces emergency electricity usage by nearly $30\%$, and lowers feed in power export by $15$–$17\%$. Moreover, it achieves a substantially higher storage level—more than three times that of the baseline mechanisms (but still in a rational level)—demonstrating more effective ESS utilization and enhanced operational stability. These improvements confirm that JPQ provides a more balanced, low-carbon and economically efficient market clearing policy for multi-microgrid energy trading.
  \begin{table}[htbp]
    \centering
    \caption{Average Hourly Performance Comparison Under Different Market Clearing Mechanisms}
    \renewcommand{\arraystretch}{1.15}
    \setlength{\tabcolsep}{3pt}
    \begin{tabular}{lcccc}
    \hline
    \textbf{Clearing} & \textbf{Reward} & \textbf{Emergency Bought} & \textbf{Feed In Sold} & \textbf{Storage} \\
    \textbf{Mechanism}  & (\$) & (kWh) & (kWh) & (kWh) \\
    \hline
    Greedy & -5.87 & 3.37 & 7.11 & 14.87 \\
    MRDA   & -5.92 & 3.38 & 7.30 & 11.95 \\
    VVDA   & -5.94 & 3.42 & 7.21 & 10.13 \\
    \hline
    JPQ   & \textbf{-3.98} & \textbf{2.40} & \textbf{6.10} & \textbf{38.04} \\
    \hline
    \end{tabular}
    \label{tab:market_mechanism_comparison}
   \end{table}

  In addition, we also analyze the performance of individual microgrids under different market mechanisms based on the result shown in Table~\ref{tab:market_mechanism_comparison}.
  \begin{table}[htbp]
  \centering
  \caption{Average Hourly Behavior of Individual Microgrids Under Different Market Clearing Mechanisms}
  \renewcommand{\arraystretch}{1.15}
  \setlength{\tabcolsep}{3pt}
  \begin{tabular}{llcccc}
    \hline
    \textbf{Grid} & \textbf{Mechanism} & \textbf{Reward} & \textbf{Emergency Bought} & \textbf{Feed In Sold} & \textbf{Storage} \\
                 &                    & (\$)            & (kWh)                     & (kWh)                 & (kWh)            \\
    \hline
    \multirow{4}{*}{Grid 1}
      & JPQ    & \textbf{-2.02} & 0.91 & 0.33 & 2.59 \\
      & Greedy & -3.30 & \textbf{0.63} & 1.56 & 1.84 \\
      & MRDA   & -2.78 & 1.24 & 0.80 & 0.99 \\
      & VDA    & -3.28 & 1.58 & 0.74 & 0.89 \\
    \hline
    \multirow{4}{*}{Grid 2}
      & JPQ    &  \textbf{0.45} & \textbf{0.03} & 0.89 & 8.04 \\
      & Greedy &  0.37 & 0.26 & 0.61 & 5.21 \\
      & MRDA   & -0.10 & 0.17 & 1.28 & 2.01 \\
      & VDA    & -0.13 & 0.14 & 1.39 & 2.13 \\
    \hline
    \multirow{4}{*}{Grid 3}
      & JPQ    & -3.43 & \textbf{1.42} & 2.06 & 2.76 \\
      & Greedy & -3.68 & 2.08 & 1.88 & 2.45 \\
      & MRDA   & -3.15 & 1.72 & 1.99 & 2.71 \\
      & VDA    & \textbf{-2.93} & 1.65 & 1.66 & 5.30 \\
    \hline
    \multirow{4}{*}{Grid 4}
      & JPQ    &  \textbf{1.10} & \textbf{0.01} & 2.82 & 24.52 \\
      & Greedy &  0.65 & 0.461 & 3.05 & 5.36 \\
      & MRDA   &  0.31 & 0.154 & 3.22 & 6.60 \\
      & VDA    & -0.10 & 0.288 & 3.50 & 0.42 \\
    \hline
  \end{tabular}
  \label{tab:agent_behavior}
    \end{table}
    Table~\ref{tab:agent_behavior} reports the individual performance of four microgrids under different market mechanisms. For deficit-type microgrids, JPQ yields the lowest or near-lowest emergency purchases (e.g., 0.91 kWh for Grid 1 and 1.42 kWh for Grid 3), indicating improved intraday balance and reduced dependence on external grid support. For surplus-type microgrids, JPQ promotes risk-aware energy reservation rather than aggressive export. This is reflected in substantially higher storage utilization under JPQ, most notably for Grid 4 (24.52 kWh under JPQ versus 0.42–6.60 kWh under the baselines) and Grid 2 (8.04 kWh versus 2.01–5.21 kWh). These patterns suggest that microgrids under JPQ strategically trade short-term feed-in opportunities for enhanced flexibility against future uncertainty, which not only contributes to lower emergency demand at the system level but also leads to more stable individual profit outcomes.

  \subsection{Performance Comparison of LSTM-MAPPO}
\begin{figure}
    \centering
    \includegraphics[width=0.48\textwidth]{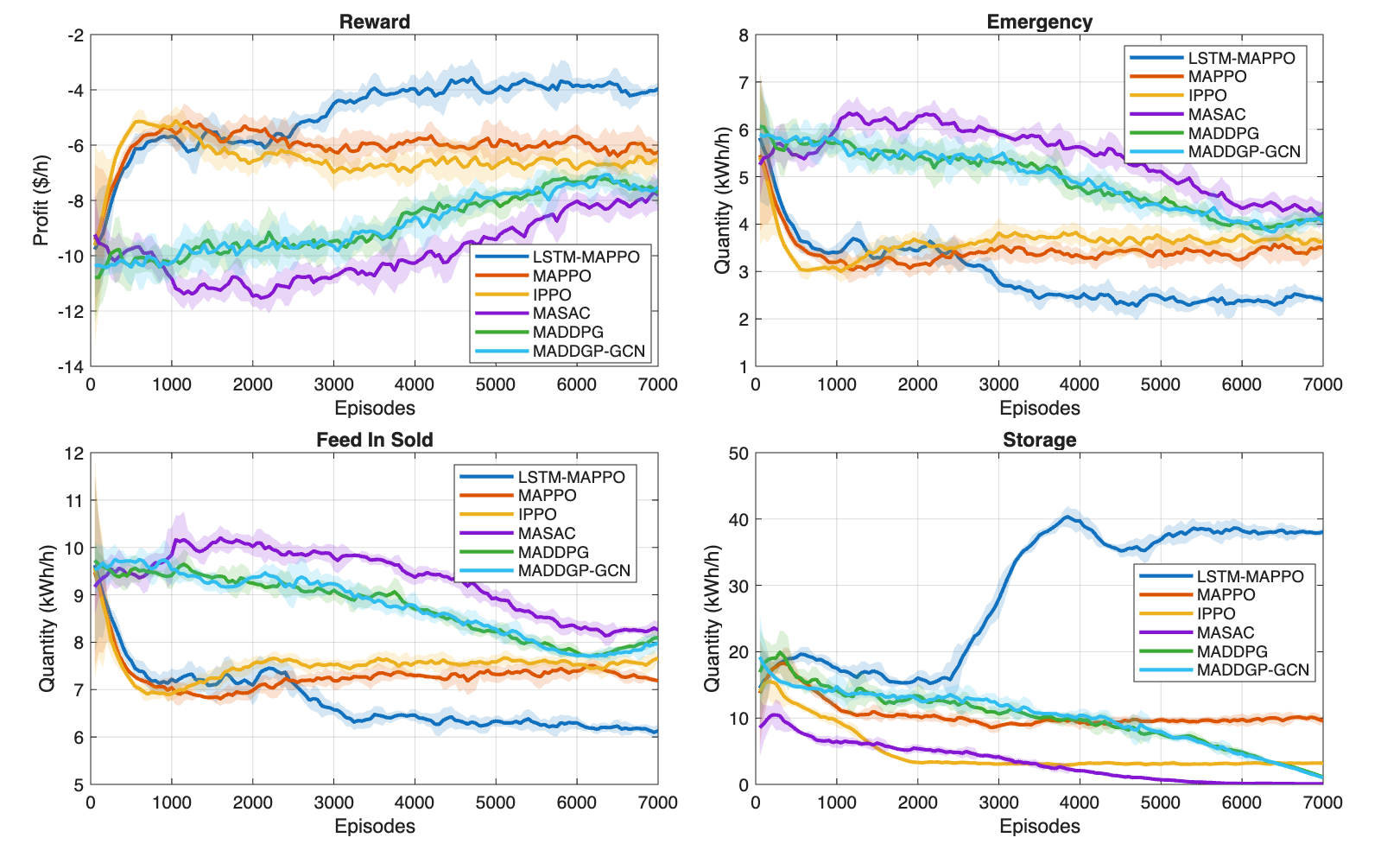}
    \caption{Training performance of six MARL algorithms over 7000 episodes. Each
    point denotes the hourly average across the four microgrids in a simulation trajectory.}
    \label{fig:algorithm_comparison}
  \end{figure}

  The performance comparison for the multi-microgrid system with different MARL
  algorithms is illustrated in Fig.~\ref{fig:algorithm_comparison} and Table~\ref{tab:marl_comparison}.
  These results show that the proposed LSTM-MAPPO algorithm consistently
  outperforms all baseline MARL algorithms across all four metrics. LSTM-MAPPO achieves
  the highest total reward, lowest emergency electricity demand, lowest feed in power
  quantity, and significantly higher utilization of the ESS. Relative to MAPPO (outperform among baselines),
  LSTM-MAPPO improves economic performance by $36.7\%$, decreases emergency electricity
  usage by $31.3\%$, reduces feed in power export by $15.1\%$, and enhances ESS utilization
  by nearly $300\%$. These consistent improvements across all metrics confirm
  that incorporating temporal dependencies through the LSTM encoder enables more
  foresighted policies, leading to higher system reliability, improved
  environmental performance, and superior overall economic welfare. In general, policy-based
  MARL algorithms (LSTM-MAPPO, MAPPO, IPPO) significantly outperform value-based or mixed-strategy
  methods (MASAC, MADDPG, MADDPG-GCN) in the multi-agent non-stationary competitive
  environment, while LSTM encoder also contributes to performance improvements
  significantly.
  \begin{table}[htbp]
    \centering
    \caption{Average Hourly Performance Comparison of the Multi-microgrid System with
    Different MARL Algorithms}
    \renewcommand{\arraystretch}{1.15}
    \setlength{\tabcolsep}{3pt}
    \begin{tabular}{lcccc}
      \hline
      \textbf{Algorithm} & \textbf{Reward} & \textbf{Emergency Bought} & \textbf{Feed In Sold} & \textbf{Storage} \\
                         & (\$)            & (kWh)                     & (kWh)             & (kWh)            \\
      \hline
      MAPPO              & -6.26           & 3.49                      & 7.19              & 9.59             \\
      IPPO               & -6.54           & 3.63                      & 7.66              & 3.22             \\
      MASAC              & -7.81           & 4.22                      & 8.26              & 0.10             \\
      MADDPG             & -7.73           & 4.18                      & 8.09              & 1.13             \\
      MADDPG-GCN         & -7.52           & 4.06                      & 7.97              & 1.02             \\
      \hline
      LSTM-MAPPO         & \textbf{-3.96}  & \textbf{2.40}             & \textbf{6.10}     & \textbf{38.04}   \\
      \hline
    \end{tabular}
    \label{tab:marl_comparison}
  \end{table}

  \section{Conclusion}
  This paper studied the intraday P2P electricity trading problem in multi-microgrid systems under renewable generation and demand uncertainties. To address the limitations of existing coordination approaches, we developed a decentralized market-learning framework that integrates market clearing mechanisms with sequential decision-making by self-interested microgrids operating under partial observability. The resulting problem was formulated as a DEC-POMDP, allowing individual microgrids to autonomously adapt their bidding and energy storage strategies in response to evolving market conditions. A carbon-aware JPQ market clearing mechanism was further introduced to shape individual incentives and trading priorities without imposing rigid coordination or emission constraints. By jointly leveraging pricing signals and double-auction clearing rules, the proposed mechanism encourages local energy balancing and risk-aware storage utilization, thereby reducing reliance on high-carbon emergency power. This work highlights the potential of combining incentive-aware market mechanisms with decentralized learning to support scalable, low-carbon operation in future multi-microgrid systems, and provides a flexible foundation for extending P2P market designs to more complex energy communities.
  Future work will extend the framework to incorporate day-ahead procurement decisions and address privacy-preserving policy learning for microgrids.



  \section*{Acknowledgment}
  Parts of this article have been grammatically revised using ChatGPT \cite{openai2023chatgpt} to improve readability. The code accompanying this paper is publicly available at \url{https://github.com/ryan-ntu/P2P_Electricity_Market_JPQ_Auction}.

  \ifCLASSOPTIONcaptionsoff
  \newpage
  \fi

  \bibliographystyle{IEEEtran}
  \bibliography{IEEEabrv,ref}
\end{document}